\documentclass[]{article}
\usepackage{amsfonts}
\usepackage{theorem}
\newtheorem{definition}{Definition}
\newtheorem{theorem}{Theorem}
\newtheorem{corollary}{Corollary}
\newtheorem{lemma}{Lemma}

\newcommand{\openbox}{$\begin{array}{c}
\hspace*{-0.55em}\sqcap \hspace*{-0.60em}\\[-0.4em] \hline
\multicolumn{1}{c}{\hspace*{-0.60em}}\\[-0.8em]
\end{array}
$}

\usepackage{enumerate}
\usepackage[unicode,colorlinks,plainpages=false]{hyperref}

\begin{document}

\centerline{{\bf Retractable state-finite automata without outputs}\footnote{AMS Subject Classification: 68Q70; Keywords: automata, congruences on automata;
Research supported by the Hungarian NFSR grant No T029525 and  No T042481}}

\bigskip

\centerline{\bf Attila Nagy}
\centerline{Department of Algebra}
\centerline{Institute of Mathematics}
\centerline{Budapest University of Technology and Economics}
\centerline{e-mail: nagyat@math.bme.hu}

\bigskip

\begin{abstract}
A homomorphism of an automaton ${\bf A}$ without outputs onto a
subautomaton ${\bf B}$ of ${\bf A}$ is called a retract homomorphism if it
leaves the elements of $B$ fixed. An automaton ${\bf A}$ is called a retractable automaton
if, for every subautomaton ${\bf B}$ of ${\bf A}$, there is a retract
homomorphism of ${\bf A}$ onto ${\bf B}$. In \cite{1} and \cite{3}, special retractable
automata are examined. The purpose of this paper is to give a construction
for state-finite retractable automata without outputs.
\end{abstract}

\bigskip

In this paper, by an automaton we mean an automaton without outputs, that is,
a system ${\bf A}=(A,X,\delta)$ consisting of a non-empty
{\it state set} $A$, a non-empty {\it input set} $X$ and a {\it transition function}
$\delta :A\times X\mapsto A$.
If $A$ has only one element then the automaton $\bf A$ will be called
{\it trivial}. The function $\delta$ is extended to
$A\times X^*$ ($X^*$ denotes the free monoid over $X$) as follows. If $a$ is
an arbitrary state of $\bf A$ then $\delta (a,e)=a$ for the empty word $e$,
and $\delta (a,qx)=\delta (\delta (a,q),x)$ for every $q\in X^*$, $x\in X$.

If $B$ is a non-empty subset of the state-set of an automaton ${\bf A}=(A,X,\delta )$
such that $\delta (b,x)\in B$ for every $b\in B$
and $x\in X$, then ${\bf B}=(B,X,\delta _B)$ is an automaton, where $\delta _B$
denotes the restriction of $\delta$ to $B\times X$. This automaton is called a
{\it subautomaton} (more precisely, an $A$-{\it subautomaton}) of ${\bf A}$.
A subautomaton ${\bf B}$ of an automaton $\bf A$ is called a
{\it proper subautomaton} of ${\bf A}$ if $B$ is a proper
subset of $A$. A subautomaton $\bf B$ of an automaton $\bf A$
is said to be a {\it minimal subautomaton} of $\bf A$ if $\bf B$ has no
proper subautomaton. If a subautomaton $\bf B$ of an automaton $\bf A$ has only
one state then $\bf B$ is minimal; the state of $\bf B$ is called a {\it trap} of $\bf A$.
If an automaton ${\bf A}=(A,X,\delta)$
contains only one trap denoted by $a_0$ then $\bf A$ is called a {\it one-trap automaton}
(or an OT-{\it automaton}). This fact will be denoted by $(A,X,\delta ;a_0)$.
If an automaton $\bf A$ has a subautomaton
which is contained in every subautomaton of $\bf A$ then it is called the
{\it kernel} of $\bf A$. The kernel of $\bf A$ is denoted by $Ker {\bf A}$.

Let ${\bf A}=(A,X,\delta)$ be an automaton containing at most one trap.
Let $A^0$ denote the following set.
$A^0=A$ if ${\bf A}$ does not contain a trap or ${\bf A}$ is trivial;
$A^0=A-\{ a_0\}$ if ${\bf A}$ is a non-trivial OT-automaton and $a_0$ is the trap
of ${\bf A}$. Consider the mapping $\delta ^0:A^0\times X\mapsto A^0$ which is
defined for a couple $(a,x)\in A^0\times X$ if and only if
$\delta (a,x)\in A^0$. In this case, let $\delta ^0(a,x)=\delta (a,x)$.
$(A^0,X,\delta ^0)$ is a {\it partial automaton} which will be denoted by ${\bf A}^0$.

An equivalence relation $\alpha$ of the state set $A$ of an automaton
${\bf A}=(A,X,\delta)$ is called a {\it congruence} of $\bf A$ if, for every
$a,b\in A$ and $x\in X$, the assumption $(a,b)\in \alpha$ implies
$(\delta (a,x),\delta (b,x))\in \alpha$. It is easy
to see that if $\bf B$ is a subautomaton of an automaton $\bf A$
then $\rho_B=\{(a,b)\in A\times A: a=b\ \hbox{or}\ a,b\in B\}$ is a
congruence of ${\bf A}$, which is called the {\it Rees congruence} of $\bf A$ induced by $\bf B$.
The factor automaton ${\bf A}/\rho _B$ is called the {\it Rees factor automaton} of
${\bf A}$ modulo ${\bf B}$.
If $\bf B$ is a subautomaton of an automaton $\bf A$ then
we may describe the Rees factor ${\bf A}/\rho _B$ as the result of
collapsing $B$ into a trap $a_0$ of the Rees factor, while the elements of
$A$ outside of
$B$ retain their identity. Sometimes we can identify these elements $a$
($a\in A-B$) with the one-element $\rho _B$-class $[a]$, that is, we can
suppose that the state set of the Rees factor is $(A-B)\cup \{ a_0\}$.

If $a$ is a state of an automaton ${\bf A}$, then the smallest
subautomaton ${\bf R}(a)$ of ${\bf A}$
containing the state $a$ is called the {\it principal subautomaton} of ${\bf A}$
generated by $a$. It is easy to see that $R(a)=\delta (a,X^*)=
\{ \delta (a,p):\ p\in X^*\}$. Clearly, every minimal subautomaton of an
automaton is principal.

The relation $\cal R$ on an automaton ${\bf A}$ defined by
${\cal R}=\{ (a,b)\in A\times A:\ R(a)=R(b)\}$ is an equivalence relation
on $A$. The ${\cal R}$-class of $A$ containing an element $a\in A$ is denoted by $R_a$.
The subset $R(a)-R_a$ is denoted by $R[a]$. It is clear that $R[a]$ is either
empty or $(R[a],X,\delta _{R[a]})$ is a subautomaton of ${\bf A}$.
The factor automaton ${\bf R}\{ a\}={\bf R}(a)/\rho _{R[a]}$ is called a
{\it principal factor} of ${\bf A}$. We note that if ${\bf R}[a]=\emptyset$
then ${\bf R}\{ a\}$ is defined to be ${\bf R}(a)$. For example, if $a$ is a
trap then $R(a)=\{ a\}$ and so $R[a]=\emptyset$.

A mapping $\phi$ (acting on the left) of the state set $A$ of an
automaton ${\bf A}=(A,X,\delta_A)$
into the state set $B$ of an automaton ${\bf B}=(B,X,\delta_B)$ is called a
{\it homomorphism} of ${\bf A}$ into ${\bf B}$ if $\phi (\delta_A (a,x))=
\delta _B(\phi (a),x)$ for every $a\in A$ and $x\in X$.

A mapping $\phi$ (acting on the left) of $A^0$ into $B^0$ is
called a {\it partial homomorphism} of a partial automaton
${\bf A}^0=(A^0,X,\delta ^0_A)$ into a partial automaton
${\bf B}^0=(B^0,X,\delta ^0_B)$ if, for every $a\in A^0$, $x\in X$,
the assumption $\delta _A(a,x)\in A^0$ implies $\delta _B(\phi (a),x)\in B^0$
and $\delta _B(\phi (a),x)=\phi (\delta _A(a,x))$.

\begin{definition}\label{df1} A subautomaton $\bf B$ of an automaton $\bf A$ is
said to be a retract subautomaton if there is a homomorphism of
$\bf A$ onto $\bf B$ which leaves the elements of $B$ fixed. Such a
homomorphism is called a retract homomorphism of $\bf A$ onto $\bf B$.
\end{definition}

\begin{definition}\label{df2} An automaton $\bf A$ is called a retractable automaton if
every subautomaton of $\bf A$ is retract.
\end{definition}

\begin{lemma}\label{lm1} Every subautomaton of a retractable automaton is retractable.
\end{lemma}
{\bf Proof}. As a subautomaton ${\bf C}$ of a subautomaton $\bf B$ of an
automaton $\bf A$ is also a subautomaton of $\bf A$, and the retriction
of a retract homomorphism of $\bf A$ onto $\bf C$ to $\bf B$ is a retract
homomorphism of $\bf B$ onto $\bf C$, our assertion is obvious.
\hfill\openbox

\begin{lemma}\label{lm2} If ${\bf A}$ is a retractable automaton and $\{ a_i:\ i\in I\}$
are elements of $A$ such that $R(a_i)\subseteq R(b)$ for an element
$b$ of $A$ then there is an index $j\in I$
such that $R(a_i)\subseteq R(a_j)$ for every $i\in I$.
\end{lemma}
{\bf Proof}. Let ${\bf A}=(A,X,\delta)$ be a retractable automaton and
$\{ a_i:\ i\in I\}$ be arbitrary elements of $A$ such that $R(a_i)\subseteq R(b)$
for an element $b$ of $A$.
Let $R=\cup _{i\in I}R(a_i)$. As ${\bf R}=(R,X,\delta _R)$ is a subautomaton
of $\bf A$, there is a retract homomorphism
$\lambda _R$ of $\bf A$ onto $\bf R$. As $\lambda _R(b)\in R$,
there is an index $j\in I$ such that $\lambda _R(b)\in R(a_j)$.
Then $\lambda _R (\delta(b,p))=
\delta (\lambda _R(b),p)\in R(a_j)$ for every $p\in X^*$, and so
$\lambda _R(R(b))\subseteq R(a_j)$. As $R(a_i)\subseteq R\cap R(b)$ ($i\in I$),
we get $R(a_i)=\lambda _R(R(a_i))\subseteq R(a_j)$ for every $i\in I$.
\hfill\openbox

\begin{corollary}\label{cr1} Every subautomaton of a principal subautomaton of a retractable
automaton is principal. In particular, for every state $a$ of a retractable
automaton $\bf A$, $R[a]$ is either empty or ${\bf R}[a]$ is a principal subautomaton of $\bf A$.
\end{corollary}
{\bf Proof}. Let $\bf B$ be a subautomaton of a principal subautomaton ${\bf R}(b)$
of a retractable automaton $\bf A$. Then $R(a)\subseteq R(b)$ for every $a\in B$.
By Lemma~\ref{lm2}, there is an element $c\in B$ such that $R(a)\subseteq R(c)$ for
every $a\in B$. As $B=\cup _{a\in B}R(a)$, we get $B=R(c)$.
\hfill\openbox

\medskip

Let $T$ be a set with a partial ordering $\leq$ such that every
two-element subset of $T$ has a lower bound in $T$ and every non-empty
subset of $T$ having an upper bound in $T$ contains a greatest element.
Then $T$ is a semilattice under multiplication $*$ by letting
$a*b$ ($a,b\in T$) be the (necessarily unique) greatest lower bound
of $a$ and $b$ in $T$. A semilattice which can be constructed as above
is called a {\it tree} (\cite{4}).

\begin{corollary}\label{cr2} A state-finite retractable
automaton $\bf A$ contains a kernel if and only if the principal subautomata
of $\bf A$ form a tree with respect to inclusion.
\end{corollary}
{\bf Proof}. Let $\bf A$ be a state-finite retractable automaton. The inclusion
(the inclusion of the state-sets) is a partial ordering on the set $T$ of
all principal subautomata of $\bf A$. By Lemma~\ref{lm2}, every non-empty
subset of $T$ having an upper bound in $T$ contains a greatest element.
As every finite tree has a least element, $T$ (which is finite) is a tree
if and only if it has a least element. As the least element of $T$ is the
kernel of $\bf A$, our proof is complete.
\hfill\openbox

\begin{lemma}\label{lm3} Every principal subautomaton of a state-finite retractable automaton
contains exactly one minimal subautomaton.
\end{lemma}
{\bf Proof}. From the finiteness of the state set, it follows that every principal subautomaton
contains a minimal subautomaton. As a minimal subautomaton is a principal
subautomaton, our assertion follows from Lemma~\ref{lm2}.
\hfill\openbox

\begin{lemma}\label{lm4} If $a_1, a_2$  are states of a state-finite retractable
automaton ${\bf A}=(A,X,\delta)$ such that $B_1\subseteq R(a_1)$,
$B_2\subseteq R(a_2)$ for distinct minimal subautomata ${\bf B}_1$ and
${\bf B}_2$ of $\bf A$ then $R(a_1)\cap R(a_2)=\emptyset$.
\end{lemma}
{\bf Proof}. If $c\in R(a_1)\cap R(a_2)$ then, by Lemma~\ref{lm3}, there is
a minimal subautomaton $\bf B$ of $\bf A$ such that
$B\subseteq R(c)\subseteq R(a_1)\cap R(a_2)$. Using again Lemma~\ref{lm3}, we get
$B_1=B=B_2$ which is a contradiction.
\hfill\openbox

\medskip

If ${\bf A}_i=(A_i,X,\delta _i)$, $i\in I$ are automata such that
$A_i\cap A_j=\emptyset$ for every $i\neq j$, then ${\bf A}=(A,X,\delta)$
is an automaton, where $A=\cup _{i\in I}A_i$ and $\delta (a,x)=\delta _i(a,x)$
for every $a\in A_i$ and $x\in X$. The automaton $\bf A$ is called the {\it direct sum} of
the automata ${\bf A}_i$, $i\in I$.

\begin{definition}\label{df3} We say that an automaton ${\bf A}$ is a strong direct
sum of a family of subautomata ${\bf A}_i$, $i\in I$
if $\bf A$ is a direct sum of ${\bf A}_i$, $i\in I$ and,
for every couple $(i,j)\in I\times I$, there is a
homomorphism of ${\bf A}_i$ into ${\bf A}_j$.
\end{definition}

\begin{theorem}\label{th1} A strong direct sum of retractable automata is retractable.
\end{theorem}
{\bf Proof}. Assume that an automaton ${\bf A}=(A,X,\delta)$ is a
strong direct sum of automata ${\bf A}_i=(A_i,X,\delta _i)$, $i\in I$. Let $\phi _{i,j}$ be the corresponding
homomorphism of ${\bf A}_i$ into ${\bf A}_j$ ($i,j\in I$). Let ${\bf R}$ be
an arbitrary subautomaton of $\bf A$. Let $R_i=R\cap A_i$. It is clear that $R_i$ is either
empty or ${\bf R}_i=(R_i,X,\delta _{R_i})$ is a subautomaton of ${\bf A}_i$. Let $\lambda _{R_i}$
denote a retract homomorphism of ${\bf A}_i$ onto ${\bf R}_i$ if $R_i\neq \emptyset$, and let
$i_0$ denote a fixed index, for which $R_{i_0}\neq \emptyset$. We define
a mapping $\lambda _R$ of $A$ onto $R$ as follows. If $a\in A_i$ and
$R_i=\emptyset$, then let $\lambda _R(a)=\lambda _{R_{i_0}}(\phi _{i,i_0}(a))$;
if $a\in A_i$ and $R_i\neq \emptyset$, then let $\lambda _R(a)=\lambda _{R_i}(a)$.
It is clear that $\lambda _R$ mapps $A$ onto $R$ and leaves the elements of $R$ fixed.
To prove that $\lambda _R$ is a homomorphism of $\bf A$ onto $\bf R$, let $i\in I$,
$a\in A_i$, $x\in X$ be arbitrary elements. In case $R_i=\emptyset$,
$$\lambda _R(\delta(a,x))=\lambda _{R_{i_0}}(\phi _{i,i_0}(\delta _i(a,x)))=
\lambda _{R_{i_0}}(\delta _{i_0}(\phi _{i,i_0}(a),x))=$$
$$=\delta _{i_0}(\lambda _{R_{i_0}}(\phi _{i,i_0}(a)),x)=\delta (\lambda
_R(a),x),$$
and, in case $R_i\neq \emptyset$, $$\lambda _R(\delta(a,x))=\lambda _{R_i}
(\delta _i(a,x))=\delta _i(\lambda _{R_i}(a),x)=\delta (\lambda _R(a),x),$$ because
$a, \delta (a,x)\in A_i$. Hence $\lambda _R$ is a retract homomorphism of $\bf A$
onto $\bf R$. Thus the theorem is proved.
\hfill\openbox

\begin{theorem}\label{th2} For a state-finite automaton ${\bf A}=(A,X,\delta)$, the
following assertions are equivalent:
\item {(i)} $\bf A$ is retractable;
\item {(ii)} $\bf A$ is a direct sum of finite many state-finite
retractable automata containing kernels being isomorphic
to each other.
\item {(iii)} $\bf A$ is a strong direct sum of finite many state-finite
retractable automata containing kernels.
\end{theorem}
{\bf Proof}. (i) implies (ii): Assume that ${\bf A}$ is retractable. As
$\bf A$ is finite, it has a minimal subautomaton.
Let $\{ {\bf B}_i$, $i=1,2,\dots r\}$ be the set of all distinct minimal
subautomata of $\bf A$. Let $A_i=\cup _{a\in A} \{ R(a):\ B_i\subseteq R(a)\}$,
$i=1,2,\dots ,r$. It is clear that ${\bf A}_i$ is a subautomaton of $\bf A$
and ${\bf B}_i$ is the kernel of ${\bf A}_i$ for every $i=1, \dots ,r$.
By Lemma~\ref{lm3}, for every principal subautomaton ${\bf R}(a)$ of $\bf A$,
there is a unique index $i$ such that $B_i\subseteq R(a)$. Thus
$A=\cup _{i=1}^rA_i$. By Lemma~\ref{lm4}, $A_i\cap A_j=\emptyset$ for every
$i\neq j$. Hence $\bf A$ is a direct sum of the automata ${\bf A}_i$,
$i=1,\dots ,r$. By Lemma~\ref{lm1},
every automaton ${\bf A}_i$ is retractable.  Let $i,j\in \{1, 2,\dots ,r\}$
be arbitrary. As ${\bf B}_i$ is a minimal subautomaton of $\bf A$, the retract homomorphism
$\lambda _{B_i}$ of $\bf A$ onto ${\bf B}_i$ maps ${\bf B}_j$ onto
${\bf B}_i$. Thus $|B_j|\geq |B_i|$. Similarly, $|B_i|\geq |B_j|$. Thus
$|B_i|=|B_j|$ and the restriction of $\lambda _{B_j}$ to $B_i$ is an
isomorphism of ${\bf B}_i$ onto ${\bf B}_j$. Thus (ii) is satisfied.

(ii) implies (iii): Assume that
$\bf A$ is a direct sum of the state-finite retractable automata ${\bf A}_i$,
$i=1, 2, \dots ,r$ such that each of ${\bf A}_i$ contains a kernel
$\bf B_i$, and, for every $i,j\in \{ 1, 2,\dots ,r\}$, there is an isomorphism
$\phi _{i,j}$ of ${\bf B}_i$ onto ${\bf B}_j$.
It is easy to see that $\Phi _{i,j}$ defined by
$$\Phi _{i,j}(a)=\phi _{i,j}(\lambda _{B_i}(a)), \ a\in A_i$$
is a homomorphism of ${\bf A}_i$ into ${\bf A}_j$, where $\lambda _{B_i}$
denotes a retract homomorphism of ${\bf A}_i$ onto ${\bf B}_i$. Thus $\bf A$
satisfies (iii).

(iii) implies (i): By Theorem~\ref{th1}, it is obvious.
\hfill\openbox

\medskip

By the previous theorem, we concentrate our attention to state-finite retractable
automata containing a kernel. These automata will be described by Corollary~\ref{cr3}
and Theorem~\ref{th7}. First consider some results and notions which will be needed for us.

\begin{lemma}\label{lm5}
Every principal factor of an automaton can contain at most one
trap.
\end{lemma}
{\bf Proof}. If $R[a]=\emptyset$ for a state $a$ then the principal factor
${\bf R}\{ a\}$ has a trap only that case when $a$ is a trap of $\bf A$, that is,
the principal factor is trivial. If $R[a]\neq \emptyset$ then $R(b)=R(a)$ for every
$b\in R_a=R(a)-R[a]$, and so ${\bf R}\{ a\}$
contains only one trap, namely the $\rho _{R[a]}$-class $R[a]$ of ${\bf R}(a)$.
\hfill\openbox

\begin{definition}\label{df4} An automaton ${\bf A}=(A,X,\delta)$ is called strongly
connected
if, for every couple $(a,b)\in A\times A$, there is a word $p\in X^+$
($X^+$ denotes the free semigroup over $X$) such that $b=\delta (a,p)$.
\end{definition}

\medskip

We note that every strongly connected automaton can contain only one subautomaton,
namely itself. We also note that if an automaton is trivial (has only one state
which is a trap) then it is strongly connected. If an automaton has at least two
state and has a trap then it is not strongly connected.

\begin{definition}\label{df5}
A non-trivial OT-automaton ${\bf A}=(A,X,\delta;a_0)$ is called strongly trap-connected
if, for every couple $(a,b)\in A\times A$, $a\neq a_0$, there is a word
$p\in X^+$ such that $b=\delta (a,p)$.
\end{definition}

\medskip

We note that every strongly trap-connected automaton ${\bf A}=(A,X,\delta ;a_0)$
contains only two subautomaton, namely
itself and $(\{ a_0\},X,\delta _{\{ a_0\}})$. Moreover, for every state $a\neq a_0$
of $\bf A$ there is a word
$p\in X^+$ such that $a=\delta (a,p)$.

\medskip

\begin{definition}\label{df6} We say that a non-trivial OT-automaton
${\bf A}=(A,X,\delta ;a_0)$ is strongly trapped if $\delta (a,x)=a_0$
for every $a\in A$ and $x\in X$.
\end{definition}

\begin{theorem}\label{th3} Every principal factor of an
automaton is either strongly connected or strongly trap-connected or strongly trapped.
\end{theorem}
{\bf Proof}. If $R[a]=\emptyset$ then ${\bf R}\{ a\}={\bf R}(a)$
is strongly connected.
If $R[a]\neq \emptyset$ then, by Lemma~\ref{lm5}, ${\bf R}\{ a\}$ is a non-trivial
OT-automaton. Let $a_0$ denote the trap of ${\bf R}\{ a\}$.
If $|R_a|=1$, that is, $R\{ a\}=\{ a,a_0\}$, then ${\bf R}\{
a\}$ is either
strongly trapped (if $\delta (a,x)\in R[a]$ in $\bf A$, that is, $\delta (a,x)=a_0$
in ${\bf R}\{ a\}$ for every $x\in X$) or strongly trap-connected
(if $a=\delta (a,x)$ for some $x\in X$). If $|R_a|>1$ then, for every
elements $b,c$ of $R_a$, $c=\delta (b,p)$ for some $p\in X^+$. Moreover,
for every $b\in R_a$, there is a word $p\in X^+$ such that
$\delta (b,p)\in R[a]$ in $\bf A$, that is, $\delta (b,p)=a_0$ in ${\bf R}\{ a\}$.
Hence ${\bf R}\{ a\}$ is strongly trap-connected.
\hfill\openbox

\begin{definition}\label{df7} An automaton $\bf A$ is called semiconnected if every
principal factor of $\bf A$ is either strongly connected or strongly trap-connected.
\end{definition}

\begin{theorem}\label{th4}
An automaton ${\bf A}=(A,X,\delta)$ is  semiconnected if and only if every
subautomaton ${\bf B}$ of ${\bf A}$ satisfies the following: for every
$a\in B$ there are elements $b\in B$ and $p\in X^+$ such that
$a=\delta (b,p)$.
\end{theorem}
{\bf Proof}. Let ${\bf A}=(A,X,\delta)$ be a semiconnected automaton and
${\bf B}$ be a subautomaton of ${\bf A}$. Let $a$ be an arbitrary element
of $B$. Then $R(a)\subseteq B$. If $a$ is a trap then $a=\delta (a,x)$ for
every $x\in X$. Consider the case when $a$ is not a trap. Then $|R(a)|\geq 2$.
If $R[a]=\emptyset$ then, by Theorem~\ref{th3}, ${\bf R}(a)={\bf R}\{ a\}$ is strongly
connected which means that, for every $b\in R(a)$ there is a word $p\in X^+$
such that $a=\delta (b,p)$. If $R[a]\neq \emptyset$ then, by Theorem~\ref{th3},
${\bf R}\{ a\}$ is strongly trap-connected and so, for every element $b\in R_a$,
there is a word $p\in X^+$ such that $a=\delta (b,p)$. Thus, in all cases,
there is a state $b\in B$ and a word $p\in X^+$
such that $a=\delta (b,p)$.

Conversely, assume that every subautomaton of an automaton ${\bf A}$ satisfies
the condition of the theorem. We show that ${\bf A}$ is semiconnected.
Let $a$ be an arbitrary element of $A$. If $a$ is a trap of ${\bf A}$
then the principal factor ${\bf R}\{ a\}$ is trivial (and so it is strongly connected).
Consider the case when $a$ is not a trap of ${\bf A}$. Then
$a$ is an element of ${\bf R}\{ a\}$ (and is not the trap of ${\bf R}\{ a\}$).
By Theorem~\ref{th3}, it is sufficient to show that the principal factor ${\bf R}\{ a\}$ is not
strongly trapped. As ${\bf R}(a)$ is a subautomaton of ${\bf A}$, by the condition of the
theorem, there are elements $b\in R(a)$ $p\in X^*$ and $x\in X$ such that
$a=\delta (b,px)=\delta (\delta (b,p),x)$ in $\bf A$. It is clear that
$b'=\delta (b,p)\notin R[a]$ and so $a=\delta (b',x)$ in ${\bf R}\{ a\}$.
Thus ${\bf R}\{ a\}$ is not strongly trapped.
\hfill\openbox

\begin{definition}\label{df8} Let ${\bf B}=(B,X,\delta_B)$ be a subautomaton of an automaton
${\bf A}=(A,X,\delta)$. We say that $\bf A$ is a dilation of $\bf B$ if there is a mapping
$\phi$ of $A$ onto $B$ which leaves the elements of $B$ fixed and
$\delta (a,x)=\delta _B(\phi (a),x)$ for all $a\in A$ and $x\in X$.
\end{definition}

\begin{theorem}\label{th5} Every dilation of a retractable automaton is retractable.
\end{theorem}
{\bf Proof}.
Let ${\bf A}=(A,X,\delta)$ be a dilation
of a retractable subautomaton ${\bf B}=(B,X,\delta _B)$. Then there is a
mapping  $\phi$ of $A$ onto $B$ which leaves the elements of
$B$ fixed and $\delta (a,x)=\delta _B(\phi (a),x)$ for every $a\in A$
and $x\in X$. Let ${\bf R}$ be a subautomaton of ${\bf A}$. Then, for every
$c\in R$ and $x\in X$, $\delta (c,x)\in R\cap B$. Let $\lambda _{R\cap B}$
denote a retract homomorphism of ${\bf B}$ onto the subautomaton
${\bf R\cap B}$. Define a mapping $\lambda _R$ of $A$ onto $R$ as follows. Let
$\lambda _R(a)=a$ if $a\in R$, and let $\lambda _R(a)=\lambda _{R\cap B}(\phi (a))$
if $a\notin R$. We show that $\lambda _R$ is a homomorpism of $\bf A$ onto
$\bf R$. Let $a\in A$ and
$x\in X$ be arbitrary elements. If $a\in R$ then $$\delta (\lambda _R(a),x)
=\delta (a,x)=\lambda _R(\delta (a,x)).$$
Assume $a\notin R$. Then
$$\delta (\lambda _R(a),x)=\delta _B(\lambda _{R\cap B}(\phi (a)),x)=$$
$$=\lambda _{R\cap B}(\delta _B(\phi (a),x))=\lambda _R(\delta
(a,x)),$$  because $\lambda _R(a), \delta (a,x)\in B$ and the restriction of
$\lambda _R$ to $B$ equals $\lambda _{R\cap B}$.
Hence $\lambda _R$ is a homomorphism of ${\bf A}$ onto ${\bf R}$.
As $\lambda _R$ leaves the elements of $R$ fixed, it is a retract homomorphism
of $\bf A$ onto $\bf R$. Consequently, ${\bf A}$ is a retractable automaton.
\hfill\openbox

\begin{theorem}\label{th6} Every retractable automaton is a dilation of a semiconnected
retractable automaton.
\end{theorem}
{\bf Proof}. Let ${\bf A}=(A,X,\delta)$ be a retractable automaton and let
$B=\delta (A,X)$. Then ${\bf B}=(B,X,\delta _B)$ is a subautomaton of
${\bf A}$ and so there is a retract homomorphism $\phi$ of ${\bf A}$ onto
${\bf B}$. Let $a\in A$, $x\in X$ be arbitrary elements. Then
$\delta (a,x)=\phi (\delta (a,x))=\delta _B(\phi (a),x)$. Hence
${\bf A}$ is a dilation of ${\bf B}$. By Lemma~\ref{lm1}, ${\bf B}$ is retractable.
Let $\bf R$ be an arbitrary subautomaton of ${\bf B}$. If $c\in R$ is an arbitrary
element, then $c=\delta (a,x)$ for some $a\in A$ and $x\in X$. Let $\lambda _R$
denote the retract homomorphism of ${\bf A}$ onto ${\bf R}$. Then
$\lambda _R(a)\in R$ and
$$c=\lambda _R(c)=\lambda _R(\delta (a,x))=\delta (\lambda _R(a),x).$$
Thus, by Theorem~\ref{th4}, ${\bf B}$ is semiconnected.
\hfill\openbox

\begin{corollary}\label{cr3} An automaton is retractable if and only if it is a dilation of a
semiconnected retractable automaton.
\end{corollary}
{\bf Proof}. By the previous two theorems, it is evident.
\hfill\openbox

\medskip

Theorem~\ref{th2} shows that the state-finite retractable automata are exactly the
direct sums of finite many state-finite retractable automata such that
each component in a mentioned direct sum contains a kernel,
and these kernels are isomorphic with each other.
Corollary~\ref{cr3} and the remark after Theorem~\ref{th2} show that every
component in a direct sum is a dilation of a state-finite
semiconnected retractable automaton containing a kernel.
Theorem~\ref{th7} will show how we can construct
the state-finite semiconnected retractable automata containing a kernel.
These results togethet give a complete description of state-finite retractable
automata.

\medskip

{\bf Construction} Let $T$ be a finite tree (under partial ordering
$\leq$) with the least element $i_0$. Let $i\succ j$ ($i,j\in T$)
denote the fact that $i>j$ and, for every $k\in T$, $i\geq k\geq j$ implies $i=k$
or $j=k$.

Let ${\bf A}_i=(A_i,X,\delta _i)$, $i\in T$ be a family of
disjunct automata such that

(i) ${\bf A}_{i_0}$ is strongly connected and ${\bf A}_i$
is a strongly trap-connected OT-automaton for every $i\in T$ with $i\neq i_0$.

(ii) Let $\phi_{i,i}$ denote the identity mapping of ${\bf A}_i$, and
assume that, for every $i,j\in T$ with $i\succ j$, there is a partial
homomorphism $\phi _{i,j}$ of ${\bf A}^0_i$ into ${\bf A}^0_j$ such that

(iii) for every $i\succ j$ there are elements
$a\in A^0_i$ and $x\in X$ such that $\delta _i(a,x)\notin A^0_i$ and
$\delta _j(\phi _{i,j}(a),x)\in A^0_j$ .

For arbitrary elements $i,j\in T$ with $i\geq j$, define a partial
homomorphism
$\Phi _{i,j}$ of ${\bf A}^0_i$ into ${\bf A}^0_j$ as follows. $\Phi
_{i,i}=\phi _{i,i}$ and, if $i>j$ such that $i\succ k_1\succ \dots
k_n\succ j$ then let $$\Phi _{i,j}=\phi _{k_n,j}\circ \phi _{k_{n-1},k_n}
\circ \dots \circ \phi _{k_1,k_2}\circ \phi _{i,k_1}.$$
(We note that if $i\geq j\geq k$ are arbitrary elements of $T$ then
$\Phi _{i,k}=\Phi _{j,k}\circ \Phi _{i,j}$.)

Let $A=\cup _{i\in T}A^0_i$. Define a transition function
$\delta ':A\times X \mapsto A$ as follows. If $a\in A^0_i$ and $x\in X$
then let $\delta '(a,x)=\delta _{i'[a,x]}(\Phi _{i,i'[a,x]}(a),x)$, where
$i'[a,x]$ denotes the greatest element of the set
$\{ j\in T:\ \delta _j(\Phi _{i,j}(a),x)\in A^0_j\}$.

It is easy to see that ${\bf A}=(A,X,\delta ')$ is an automaton which
will be denoted by $(A_i,X,\delta _i; \phi _{i,j},T)$.

\begin{theorem}\label{th7}
A finite automaton is a semiconnected retractable automaton containing a
kernel if and only if it is isomorphic to an automaton
$(A_i,X,\delta _i;\phi _{i,j},T)$ constructed as above.
\end{theorem}
{\bf Proof}. Let ${\bf R}$ be a subautomaton of an automaton
$(A_i,X,\delta _i;\phi _{i,j},T)$. As every automaton ${\bf A}_i$
($i\in T-\{ i_0\}$) is strongly trap-connected and ${\bf A}_{i_0}$ is
strongly connected,
it follows that $R=\cup _{j\in \Gamma}A^0_j$ for some non-empty subset
$\Gamma$ of $T$. We show that $\Gamma$ is an ideal of $T$, that is,
$i\in \Gamma$ and $j\leq i$ together imply $j\in \Gamma$ for all $i,j\in T$.
Let $i$ be an arbitrary element of $T$ such that $i\in \Gamma$, $i\neq i_0$.
If $j\in T$ with $i\succ j$ then, by (iii), there are elements
$a\in A^0_i$ and $x\in X$ such that $\delta _i(a,x)\notin A^0_i$ and
$\delta _j(\phi _{i,j}(a),x)\in A^0_j$. Then $\delta '(a,x)\in A^0_j$.
Hence $A^0_j\cap R\neq
\emptyset$ which implies that $A^0_j\subseteq R$ and so $j\in \Gamma$.
This implies that $\Gamma$ is an ideal of $T$.
As $T$ is a tree,
$$\pi:\ i\mapsto \ \hbox{max}\{\gamma\in \Gamma:\ \gamma \leq
i\}$$ is a well-defined mapping of $T$ onto $\Gamma$ which leaves the elements of
$\Gamma$ fixed (in fact, $\pi$ is a retract homomorphism of the semigroup
$T$ onto the ideal $\Gamma$ of $T$ (see \cite{4})).
We define a retract homomorphism $\lambda _R$ of ${\bf A}$ onto ${\bf R}$.
For an arbitrary element $a\in A$, let $$\lambda _R(a)=\Phi _{i, \pi (i)}(a)$$
if $a\in A^0_i$. It is easy to see that $\lambda _R$ leaves the elements of
$R$ fixed. We prove that $\lambda _R$ is a homomorphism of ${\bf A}$ onto
${\bf R}$. Let $x\in X$, $a\in A^0_i$ be arbitrary elements.
Using $\delta '(a,x)=\delta _{i'[a,x]}(\Phi _{i,i'[a,x]}(a),x)\in
A^0_{i'[a,x]}$ and
the fact that $\Phi _{i'[a,x],\pi (i'[a,x])}$ is a partial homomorphism, we get
$$\lambda _R(\delta '(a,x))=
\lambda _R(\delta _{i'[a,x]}(\Phi _{i,i'[a,x]}(a),x))=$$
$$=\Phi _{i'[a,x],\pi (i'[a,x])}(\delta _{i'[a,x]}(\Phi
_{i,i'[a,x]}(a),x))=$$
$$=\delta _{\pi (i'[a,x])}(\Phi _{i,\pi (i'[a,x])}(a),x)\in A^0_{\pi
(i'[a,x])}.$$ Using $\Phi _{i,\pi (i)}(a)\in A^0_{\pi (i)}$, we have
$$\delta '(\lambda _R(a),x)=\delta '(\Phi _{i,\pi (i)}(a),x)=$$
$$=\delta _{(\pi (i))'[\Phi _{i,\pi (i)}(a),x]}(\Phi _{\pi (i),(\pi
(i))'[\Phi _{i,\pi (i)}(a),x]}
(\Phi _{i,\pi (i)}(a)),x)=$$
$$=\delta _{(\pi (i))'[\Phi _{i,\pi (i)}(a),x]}(\Phi _{i,(\pi (i))'[\Phi _{i,\pi (i)}(a),x]}(a),x)\in A^0_{(\pi (i))'[\Phi _{i,\pi (i)}(a),x]}.$$
To prove that $\lambda _R(\delta '(a,x))=\delta '(\lambda _R(a),x)$, it is
sufficient to show that
$$(\pi (i))'[\Phi _{i,\pi (i)}(a),x]=\pi (i'[a,x]).$$
First, assume $i'[a,x]\geq \pi (i)$ (and so $\pi (i'[a,x])=\pi (i)$). As
$\phi _{i'[a,x],\pi (i)}$ is a partial homomorphism of $A^0_{i'[a,x]}$ into
$A^0_{\pi (i)}$ and $\delta _{i'[a,x]}(\Phi _{i,i'[a,x]}(a),x)\in A^0_{i'[a,x]}$,
we get $$\delta _{\pi (i)}(\Phi _{i,\pi (i)}(a),x)=
\delta _{\pi (i)}(\Phi _{i'[a,x],\pi (i)}(\Phi _{i,i'[a,x]}(a)),x)=$$
$$=\Phi _{i'[a,x],\pi (i)}(\delta _{i'[a,x]}(\Phi _{i,i'[a,x]}(a),x))\in A^0_{\pi (i)}$$
and so $$(\pi (i))'[\Phi _{i,\pi (i)}(a),x]=\pi (i)=\pi (i'[a,x]).$$
Next, consider the case when  $i'[a,x]<\pi (i)$ (and so $\pi (i'[a,x])=i'[a,x]$).
If $j\in T$ with $\pi (i)\geq j>i'[a,x]$ then we have
$$\delta _j(\Phi _{\pi (i),j}(\Phi _{i,\pi (i)}(a)),x)=
\delta _j(\Phi _{i,j}(a),x)\notin A^0_j.$$ Then
$$(\pi (i))'[\Phi _{i,\pi (i)}(a),x]\leq i'[a,x].$$ As
$$\delta _{i'[a,x]}(\Phi _{\pi (i),i'[a,x]}(\Phi _{i,\pi (i)}(a)),x)=
\delta _{i'[a,x]}(\Phi _{i,i'[a,x]}(a),x)\in A^0_{i'[a,x]},$$ we get
$$(\pi (i))'[\Phi _{i,\pi (i)}(a),x]\geq i'[a,x].$$ Hence
$$(\pi (i))'[\Phi _{i,\pi (i)}(a),x]=i'[a,x]=\pi (i'[a,x]).$$
Consequently,
$(\pi (i))'[\Phi _{i,\pi (i)}(a),x]=\pi (i'[a,x])$ in both cases. Hence $\lambda _R$
is a (retract) homomorphism of ${\bf A}$ onto ${\bf R}$. Thus
${\bf A}=(A_i,X,\delta _i;\phi _{i,j},T)$ is a retractable automaton.

We show that ${\bf A}$ is semiconnected. If ${\bf R}$ is an arbitrary subautomaton of
${\bf A}$, then there is an ideal $\Gamma$ of $T$ such that
$R=\cup _{j\in \Gamma}A^0_j$ (see above). Let $a\in R$ be an arbitrary
element. Then $a\in A^0_k$ for some $k\in \Gamma$. As $A_k$ is strongly connected
or strongly trap-connected, there are elements $b\in A^0_k$ and
$p\in X^+$ such that
$a=\delta _k(b,p)=\delta '(b,p)$. By Theorem~\ref{th4}, it means that ${\bf A}$
is semiconnected. As $i_0$ is contained in every ideal of $T$, ${\bf A}_{i_0}$
is the kernel of $(A_i,X,\delta _i;\phi _{i,j},T)$.

Conversely, let ${\bf A}$ be a finite semiconnected retractable automaton
containing a kernel. Let $Prf(A)$ denote the set of all
principal factors of $\bf A$. By Corollary~\ref{cr2}, $Prf(A)$ is a (finite) tree
under partial ordering $\leq$ defined by
${\bf R}\{ a\}\leq {\bf R}\{ b\}$ if and only if $R(a)\subseteq R(b)$.
As $\bf A$ is semiconnected, the least element of $Prf(A)$ is strongly
connected, the other ones are strongly trap-connected.

Let $T$ be a set
with $|T|=|Prf(A)|$. Denote a bijection of $T$ onto $Prf(A)$ by $f$.
Define a partial ordering $\leq$ on $T$ by $i\leq j$ ($i,j\in T$) if and only if $f(i)\leq
f(j)$. Let $i_0$ denote the least element of $T$. Clearly, $T$ is a
finite tree with the least element $i_0$. For every element $i\in T$,
fix an element $a_i$ in $A$ such that $f(i)={\bf R}\{ a_i\}$. (We note that
${\bf R}\{ a_i\}={\bf R}\{ a_j\}$ iff $a_i=a_j$ iff $i=j$). As ${\bf R}\{ a_{i_0}\}$ is
strongly connected and ${\bf R}\{ a_i\}$ is strongly trap-connected if
$i\neq i_0$, condition $(i)$ of the Construction is satisfied.

Let $\lambda _{R(a_j)}$ ($j\in T$) denote a fix retract homomorphism of $\bf A$
onto
${\bf R}(a_j)$. For every $i,j\in T$ with $i\succeq j$, let $\lambda _{i,j}$ denote the
restriction of $\lambda _{R(a_j)}$ to $R(a_i)$. It is obvious that
$\lambda _{i,j}$ is a retract homomorphism of ${\bf R}(a_i)$ onto ${\bf
R}(a_j)$ for every $i\succeq j$, ($i,j\in T$). Moreover, $\lambda _{i,i}$ is the identity mapping of
${\bf R}(a_i)$, for every $i\in T$. We show that $\lambda _{i,j}$ maps
$R_{a_i}$ into $R_{a_j}$. Let $a\in R_{a_i}$ be an arbitrary element (so
$R(a)=R(a_i)$). Then, for every $p\in X^*$, $\lambda _{i,j}(\delta
(a,p))=\delta (\lambda _{i,j}(a),p)$. If $\lambda _{i,j}(a)$ was in
$R[a_j]$ then we would have $\lambda _{i,j}(\delta (a,p))\in R[a_j]$ for every
$p\in X^*$,
because ${\bf R}[a_j]$ is a subautomaton of $\bf A$. This would imply that
$\lambda _{i,j}(R(a_i))\subseteq R[a_j]$ which is impossible, because
$\lambda _{i,j}$ maps $R(a_i)$ onto $R(a_j)=R_{a_j}\cup R[a_j]\supset
R[a_j]$. Hence $\lambda _{i,j}$ maps $R_{a_i}$ into $R_{a_j}$ and so
$\lambda _{i,j}$ can be considered as a mapping of $R^0\{ a_i\}$
into $R^0\{ a_j\}$. If $\delta (a,x)\in R_{a_i}$ for some $a\in
R_{a_i}$ and $x\in X$ then $\delta (\lambda _{i,j}(a),x)=\lambda
_{i,j}(\delta (a,x))\in R_{a_j}$. Hence $\lambda _{i,j}$ is a partial
homomorphism of the partial automaton ${\bf R}^0\{ a_i\}$ into the
partial automaton ${\bf R}^0\{ a_j\}$. Thus condition (ii) of the
Construction is satisfied (for ${\bf A}_i={\bf R}\{ a_i\}$, $\phi
_{i,j}=\lambda _{i,j}$).

Assume $i\succ j$.
Let $b\in R_{a_j}$ be an arbitrary element. Then $a_i\neq b\in R(a_i)$ and so
there is a word $p=x_1x_2\dots x_n\in X^+$ $(x_1,x_2,\dots x_n\in
X$) such that $b=\delta (a_i,p)$. Let $m$ be the least index such that
$\delta (a_i,x_1\dots x_m)\in R_{a_j}$. Consider an element $a$ of
$R_{a_i}$ (or of ${\bf R}^0\{ a_i\}$) as follows. Let $a=a_i$ if $m=1$.
Let $a=\delta (a_i,x_1\dots x_{m-1})$ if $m>1$. Then $\delta
(a,x_m)\notin R_{a_i}$ (or $\delta (a,x_m)\notin {\bf R}^0\{ a_i\}$). On
the other hand, $$\delta (\lambda _{i,j}(a),x_m)=\lambda _{i,j}(\delta
(a,x_m))=\delta (a,x_m)\in R_{a_j}=R^0\{ a_j\},$$ because $\lambda _{i,j}$
leaves the elements of $R(a_j)$ fixed. Thus (iii) of the Construction is
satisfied (for $\phi _{i,j}=\lambda _{i,j}$, $x=x_m$).

For arbitrary elements $i,j\in T$ with $i\geq j$, define the mapping
$\Phi _{i,j}$ as follows. Let $\Phi _{i,i}=\lambda _{i,i}$ and, if $i>j$
with $i\succ k_1\succ k_2\succ \dots k_n\succ j$ then let $$\Phi
_{i,j}=\lambda _{k_n,j}\circ \dots \circ \lambda _{i,k_1}.$$ It is clear
that $\Phi _{i,j}$ is a retract homomorphism of ${\bf R}(a_i)$ onto
${\bf R}(a_j)$ such that it maps $R_{a_i}$ into $R_{a_j}$. Thus $\Phi
_{i,j}$ can be considered as a partial homomorphism of ${\bf R}^0\{
a_i\}$
into ${\bf R}^0\{ a_j\}$. Moreover, $\Phi _{i,k}=\Phi _{j,k}\circ \Phi
_{i,j}$ for every $i,j,k\in T$ with $i\geq j\geq k$.

Construct the automaton
${\bf R}=(R\{ a_i\},X,\delta _i;\lambda _{i,j},T)$, where $\delta _i$ is
the transitive function of the factor automaton ${\bf R}\{ a_i\}$ induced by
$\delta$. It is clear that the state sets of the automata $\bf R$ and
$\bf A$ are the same. We show that the
transitive functions $\delta$ of ${\bf A}$ equals the transitive function
$\delta '$ of ${\bf R}$. Let $i\in T$, $a\in R_{a_i}=R^0\{ a_i\}$, $x\in
X$ be arbitrary elements. Assume $\delta (a,x)\in R_{a_j}$ ($i\geq j$).
Let $k\in T$ with $i\geq k>j$. Then $\delta (a,x)\in R[a_k]\subset R(a_k)$ and so
$$\delta (\Phi _{i,k}(a),x)=\Phi _{i,k}(\delta (a,x))=\delta (a,x)\notin R_{a_k}=R^0\{ a_k\},$$
because $\Phi _{i,k}$ leaves the elements of $R(a_k)$ fixed.
If $j\geq k$ then $$\delta (\Phi _{i,k}(a),x)=\Phi _{i,k}(\delta (a,x))=$$
$$=\Phi _{j,k}\circ \Phi _{i,j}(\delta (a,x))=\Phi _{j,k}(\delta (a,x))
\in R_{a_k}=R^0\{ a_k\},$$ because
$\Phi _{i,j}$ leaves the element $\delta (a,x)\in R_{a_j}=R^0\{ a_j\}$ fixed, and
$\Phi _{j,k}$ maps $R_{a_j}$ into $R_{a_k}$.
Consequently $i'[a,x]=j$. Hence $$\delta (a,x)=\Phi _{i,j}(\delta (a,x))=
\delta (\Phi _{i,j}(a),x)=\delta _j(\Phi _{i,j}(a),x)=$$
$$=\delta _{i'[a,x]}(\Phi _{i,i'[a,x]}(a),x)= \delta '(a,x).$$
Thus the theorem is proved.
\hfill\openbox

\end{document}